\documentclass[11pt, letterpaper]{article}

\usepackage{graphicx} 
\usepackage{amsmath,amsfonts,mathtools}
\usepackage[makeroom]{cancel}
\usepackage{float}
\usepackage[margin=1in]{geometry}
\usepackage{booktabs}
\usepackage{tabularx,color}
\usepackage[utf8]{inputenc}
\usepackage[english]{babel}
\usepackage{natbib}
\usepackage{setspace}
\usepackage{xparse}
\usepackage{multirow, makecell}
\usepackage{subcaption}
\usepackage{url}
\usepackage{caption}
\usepackage{xr}
\usepackage{soul}
\usepackage{xcolor}
\usepackage[normalem]{ulem}
\setstcolor{blue}
\usepackage{changepage}
\usepackage{subfiles}

\NewDocumentCommand{\chRA}{o o o o o}{%
  \Delta_{\IfNoValueTF{#1}{v}{#1}} \text{RA}_{\IfNoValueTF{#2}{j}{#2}\IfNoValueTF{#3}{p}{#3}}[\IfNoValueTF{#4}{t}{#4}, \IfNoValueTF{#5}{\mathbf{x}(t)}{#5}]%
}
\NewDocumentCommand{\RA}{o o o}{%
  \text{RA}_{\IfNoValueTF{#1}{j}{#1}}[\IfNoValueTF{#2}{t}{#2}, \IfNoValueTF{#3}{\mathbf{x}(t)}{#3}]%
}
\NewDocumentCommand{\chDiv}{o o o o}{%
  \Delta_{\IfNoValueTF{#1}{v}{#1}} \alpha_{\IfNoValueTF{#2}{p}{#2}}[\IfNoValueTF{#3}{t}{#3}, \IfNoValueTF{#4}{\mathbf{x}(t)}{#4}]%
}

\NewDocumentCommand{\Div}{o o o}{%
  \alpha_{\IfNoValueTF{#1}{p}{#1}}[\IfNoValueTF{#2}{t}{#2}, \IfNoValueTF{#3}{\mathbf{x}(t)}{#3}]%
}

\NewDocumentCommand{\sth}{}{%
  ^\text{th}%
}

\NewDocumentCommand{\chRAin}{o o o o o}{%
  \Delta_{\IfNoValueTF{#1}{v}{#1}} \text{RA}_{\IfNoValueTF{#2}{j}{#2}\IfNoValueTF{#3}{p}{#3}}\left\{\IfNoValueTF{#4}{t}{#4}, \IfNoValueTF{#5}{\mathbf{x}(t)}{#5}\right\}%
}
\NewDocumentCommand{\RAin}{o o o}{%
  \text{RA}_{\IfNoValueTF{#1}{j}{#1}}\left\{\IfNoValueTF{#2}{t}{#2}, \IfNoValueTF{#3}{\mathbf{x}(t)}{#3}\right\}%
}
\NewDocumentCommand{\chDivIn}{o o o o}{%
  \Delta_{\IfNoValueTF{#1}{v}{#1}} \alpha_{\IfNoValueTF{#2}{p}{#2}}\left\{\IfNoValueTF{#3}{t}{#3}, \IfNoValueTF{#4}{\mathbf{x}(t)}{#4}\right\}%
}

\NewDocumentCommand{\DivIn}{o o o}{%
  \alpha_{\IfNoValueTF{#1}{p}{#1}}\left\{\IfNoValueTF{#2}{t}{#2}, \IfNoValueTF{#3}{\mathbf{x}(t)}{#3}\right\}%
}
 
\title{A Bayesian Functional Concurrent Zero-Inflated Dirichlet-Multinomial Regression Model with Application to Infant Microbiome}

\author{
Brody Erlandson$^{1}$\thanks{Brody.Erlandson@colostate.edu}
\and
Ander Wilson$^{1}$\thanks{Ander.Wilson@colostate.edu}
\and
Matthew D. Koslovsky$^{1}$\thanks{Matt.Koslovsky@colostate.edu}
}

\date{$^{1}$Department of Statistics, Colorado State University}

\begin{document}

\maketitle

\begin{abstract}
\begin{adjustwidth}{0.25in}{0.25in}
The infant microbiome undergoes rapid changes in composition over time and is associated with long-term risks of conditions such as immune strength, allergy, asthma, and other health outcomes. Modeling the associations between exposures or treatments and microbial composition over time is essential for understanding the factors that drive these changes. Estimating these temporal dynamics has several challenges including: repeated measures, overdispersion, compositionality, high-dimensional parameter spaces, and zero-inflation. Many longitudinal regression models used in human microbiome research assume constant effects over time that cannot capture time-varying or functional effects of exposures, ignore the compositional structure of the data by modeling each taxon separately, and are not equipped to handle potential zero-inflation. Dirichlet-multinomial (DM) regression models inherently accommodate overdispersion and the compositional structure of the data and have been extended to account for excess zeros. However, existing DM-based regression models are unable to additionally handle repeated measures designs. To fill this gap, we propose a functional concurrent zero-inflated Dirichlet-multinomial (FunC-ZIDM) regression model which is designed to model time-varying relations between observed covariates and microbial taxa while accounting for zero-inflation, compositionality, and repeated measures. Through simulation, we demonstrate that the model can accurately estimate the underlying functional relations and scale to large compositional spaces. We apply our model to investigate time-varying associations between infant microbiome composition and observed covariates during the 11-week postnatal period. We found that $\alpha$-diversity \textcolor{black}{(i.e., the diversity of the microbiome within an individual)} is positively associated with a higher gestational age and percentage of breast milk in the diet. We provide an accompanying \texttt{R} package and \texttt{shiny} app to implement the method and generate plots.
\end{adjustwidth}
\end{abstract}

\begin{center}
\begin{adjustwidth}{0.25in}{0.25in}
\end{adjustwidth}
\end{center}

\section{Introduction}\label{sec:intro}

Microbial composition development in the first years of life is associated with immune health, allergy, asthma, and other health outcomes in infancy and throughout the life course \citep{gregory2016influence, durack2019gut}. While the adult microbiome remains relatively stable over time, the infant microbiome undergoes rapid, dramatic compositional shifts, which slow significantly by age 3-5 \citep{gilbert2018current, durack2019gut}. Infants born at earlier gestational ages are more likely to have poor microbial development due to underdeveloped gut barriers and immune systems, resulting in a heightened risk of negative health outcomes \citep{healy2022clinical,chernikova2018premature}. Many factors are associated with microbial composition in infancy, such as child sex, mode of delivery, maternal and infant antibiotic use, pet exposure, breastfeeding, air pollution, and maternal health \citep{durack2019gut,patterson2021prenatal,chen2021sex,valeri2021biological}. For example, studies have shown that breast milk composition and intake are associated with higher microbial diversity, potentially serving as an intervention strategy for at-risk infants \citep{cong2016gut,gregory2016influence,pannaraj2017association,ma2020comparison}. Understanding how factors may influence the initial colonization and development of infant microbiome is important for lowering the risk of negative health outcomes in childhood and adulthood.

Microbiome data pose several analytical challenges. The sequencing of the sample yields a fixed number of reads, meaning the counts represent relative, rather than absolute, abundances. This results in compositional data, where the taxa counts must sum to a fixed total. Ignoring compositionality can result in incorrect findings \citep{gloor2017microbiome}. Microbiome data are typically high-dimensional with hundreds or even thousands of different taxa in a sample. Additionally, microbiome data are overdispersed due to large within- and between-subject variation \citep{consortium2012structure, lyu2023methodological}. Finally, the data are often zero-inflated. Zero counts may occur in several ways, including the host truly having zero occurrences of a microbe (structural zero), or the microbe is present in the host but was not collected in the sample (at-risk zero). 

Many analytical methods have been proposed to model microbiome data. Commonly, researchers apply generalized linear regression models to analyze each taxon separately, such as negative binomial models which ignore potential zero-inflation and the compositional structure of the data. Zero-inflated regression models were subsequently developed and applied to handle potential zero-inflation \citep{jiang2021bayesian}. In contrast, multinomial logistic and Dirichlet-multinomial regression models were introduced to handle the compositionality, overdispersion, and high-dimensionality in the data, but do not handle excess zeros \citep{kwak2002multinomial,xia2013logistic}. Recently, \cite{koslovsky2023bayesian} developed a zero-inflated Dirichlet-multinomial (ZIDM) model and \cite{tang2019zero} developed a zero-inflated generalized DM regression model, which extend the traditional DM model to handle zero-inflation.

Microbiome data are often collected longitudinally to capture the dynamic nature of the microbiome \citep{gilbert2018current,kodikara2022statistical,lyu2023methodological}. Mixed models, such as negative binomial, zero-inflated negative binomial, zero-inflated Poisson-gamma, and Gaussian models, have been applied to investigate longitudinal microbiome data \citep{zhang2017negative,zhang2020fast,zhang2020zero,jiang2023flexible}. \cite{ridenhour2017modeling} and \cite{chen2017high} employed autoregressive and state-space models to handle the temporal structure. Alternatively in longitudinal studies, the outcome and covariates can be thought of as functions of time. Functional data analysis methods are a popular class of methods that model functional covariates and/or outcomes, resulting in scalar-on-function, function-on-scalar, or function-on-function regression models \citep{morris2015functional}. A special case of function-on-function regression is when the functional outcome and functional covariates are collected simultaneously, which is often referred to as functional concurrent or varying-coefficient regression \citep{ramsay2005functional,hastie1993varying}. These models allow the effect of a covariate to vary as a function of another variable, often time. In longitudinal data, time-varying coefficients are frequently used to capture non-constant effects. Time-varying effects have recently been introduced in zero-inflated Poisson autoregressive  \citep{mao2024zero} and zero-inflated negative binomial models \citep{piulachs2021bayesian}. However, these models ignore the compositional structure of microbiome data, which could lead to false positives due to the interdependent relations among taxa \citep{dai2019batch}.

To overcome these challenges, we propose a functional concurrent zero-inflated Dirichlet-multinomial (FunC-ZIDM) regression model. Our approach models the microbiome data as compositional, while handling zero-inflation, repeated measures, and time-varying effects. Our proposed model is scalable and estimates smooth functional effects over time through regularization. We apply the FunC-ZIDM regression model to explore the potential functional relations between infant microbial composition and a set of observed covariates. We observe functional relations between multiple covariates, including gestational age and percentage of breast milk in the infant diet with the taxa relative abundances and compositional diversity. We provide an accompanying \texttt{R} package, \texttt{FunCZIDM}, to implement the method and code to reproduce the simulation and data analysis.

\section{Preterm Infant Microbiome Data}\label{sec:data}

This work was motivated by microbial composition data collected on preterm infants by \cite{la2014patterned}. Fecal microbiome data were collected from 58 premature infants between 1 and 80 days after birth, each having 3 to 41 samples, with a median of 15 and a total of 922 samples. Most samples were collected between 5 and 50 days of life. The microbial profiles were obtained at the class level using 16S rRNA genes from infant stool samples, resulting in 29 different taxa available for analysis. In practice, taxa are typically dropped from an analysis based on the proportion of zero counts observed in the sample. For this analysis, we excluded taxa that have less than 5 individuals with at least one non-zero count as our model accommodates both zero-inflation and repeated measures data. The final dataset included 16 taxa, which had a maximum of 98.6\% zeros for a taxon. The data contained six covariates: percentage of breast milk in the diet (split into $<$$10\%$, $10-50\%$, and $>50\%$), gestational age at birth, proportion of observed days with antibiotic use, mode of delivery (vaginal or cesarean section), room type (multi-patient or single), and infant sex (male or female). Proportion of observed days with antibiotic use was the only covariate that was time-varying. 

\begin{figure}[h!]
    \centering
    \begin{subfigure}[t]{0.49\textwidth}
      \centering
        \includegraphics[width=1\linewidth]{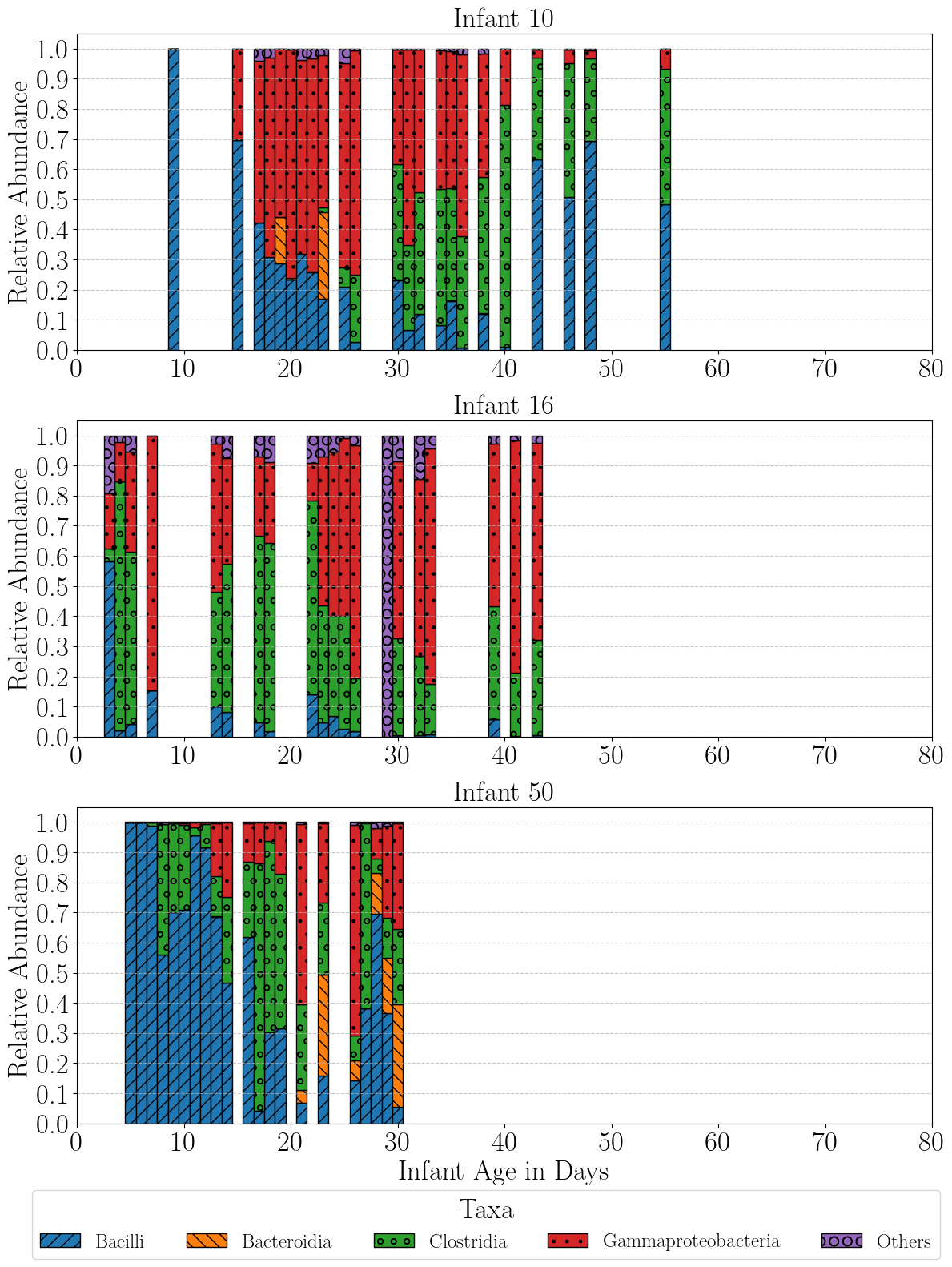} 
        \caption{Relative Abundance for Three Individual Infants} \label{fig:idvInf}
    \end{subfigure}
    \begin{subfigure}[t]{0.49\textwidth}
      \centering
        \includegraphics[width=1\linewidth]{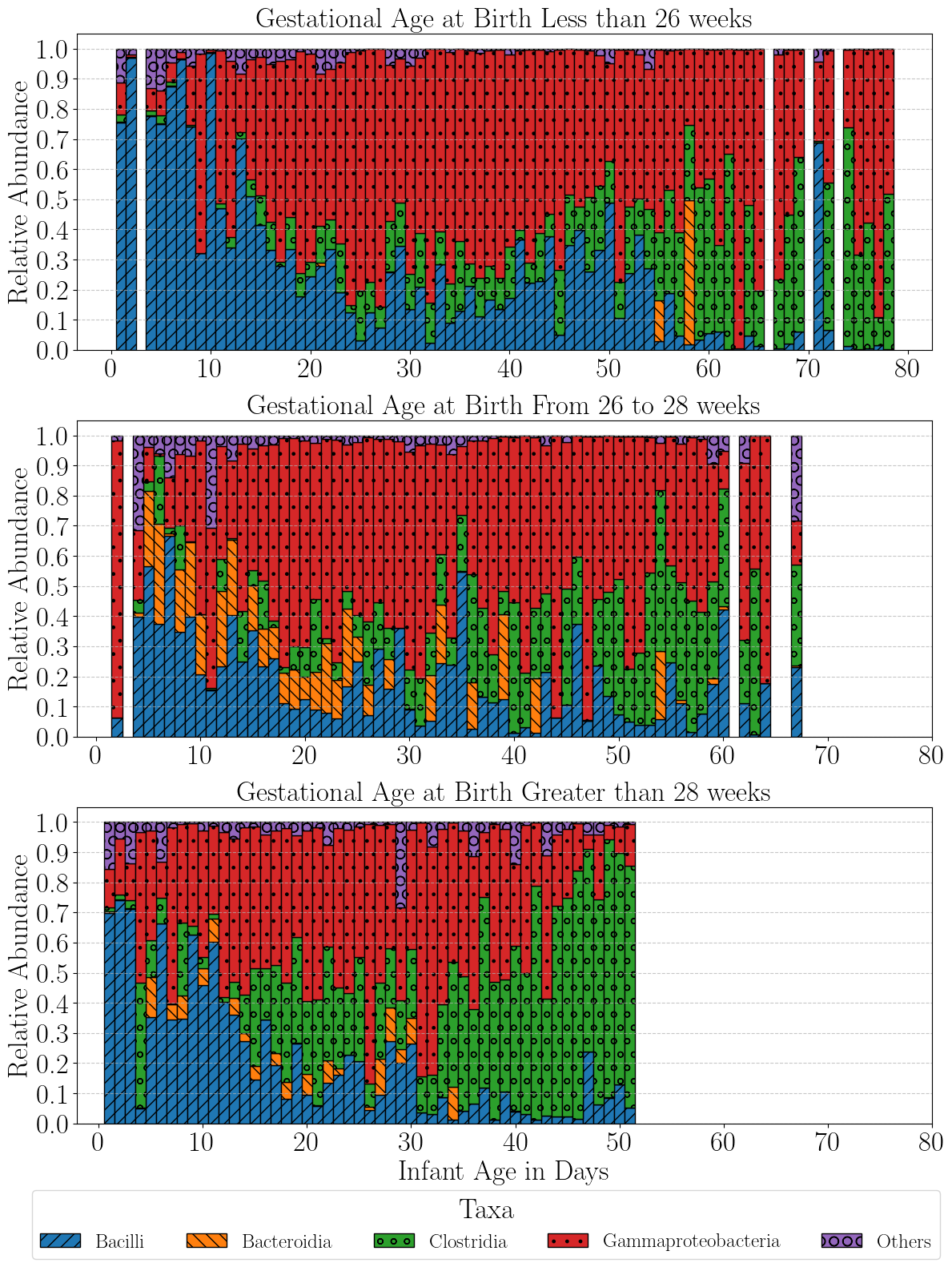} 
        \caption{Average Relative Abundance in the Sample by Age} \label{fig:infGARA}
    \end{subfigure}
    
    \caption{Observed relative abundance over time (days) for individual infants and averaged over gestational age at birth. Each plot on the left-hand side represents an individual infant's observed microbial composition on the day sampled from \cite{la2014patterned}, where the stacked bar represents the taxa relative abundances observed from that sample. The right-hand side splits gestational age at birth into three groups, and the daily average compositions are plotted as a stacked bar.}
    \label{fig:idvInfandGARA}
\end{figure}

Figure \ref{fig:idvInf} highlights the substantial variation in sampling patterns between individuals as well as variation in the microbial composition within and between individuals over time. Figure \ref{fig:infGARA} presents the average compositional shift in the microbiome over time, stratified by gestational age at birth. We observed a distinct transition for each level of gestational age at birth, suggesting an association with microbial composition. Our specific inferential goals are to understand how the relative abundance of each taxon changes over time, how covariates are associated with the relative abundances as well as how the effects may change over time, and understand how covariates impact overall microbial diversity.  

\section{Methods}\label{sec:methods}

Our motivation is to investigate the relation between $P$, potentially functional, covariates and the functional multivariate compositional counts. For individual $i = 1, \dots, N$, let $\mathbf{z}_i(t) = [z_{i1}(t),\dots,z_{iJ}(t)]^\prime$ denote the functional counts observed at time $t$ for the $J$ taxa, and $\dot{z}_i(t) = \sum_{j=1}^J z_{ij}(t)$ represent the total counts observed across all $J$ taxa for individual $i$ at time $t$. For $t\in \mathcal{T}$, where $\mathcal{T}$ can be any continuous subset of the positive reals, let $x_{ip}(t)$ represent the $p\sth$ covariate for individual $i$ observed at time $t$. Each observation can occur at any time $t$ in this space.

\subsection{A Functional Concurrent Zero-Inflated Dirichlet-Multinomial Model}\label{sec:FuncReg}

We model the observed counts as
\begin{equation*}
\begin{aligned}
    \mathbf{z}_{i}(t) &\sim \text{Multinomial}[\dot{z}_{i}(t),\boldsymbol{\psi}_i(t)],\\
\end{aligned}
\end{equation*}
where $\boldsymbol{\psi}_i(t)=[\psi_{i1}(t),\dots,\psi_{iJ}(t)]^\prime$ is a $J$-dimensional vector of relative abundances at time $t$ that are constrained to be positive and sum to 1. \textcolor{black}{To accommodate zero-inflation and enhance scalability in high-dimensional settings, we incorporate auxiliary parameters and reparameterize the Dirichlet distribution with zero-inflated gamma random variables as in \cite{koslovsky2023bayesian}, which is inspired by techniques used in Bayesian non-parametrics \citep{james2009posterior,argiento2015priori}.} Specifically, we assume

$$
\begin{aligned}
   \boldsymbol{\psi}_i(t) &= \mathbf{c}_{i}(t)/T_{i}(t) \text{ and }\\
   c_{ij}(t)|\eta_{ij} &\sim (1-\eta_{ij})\delta_0 + \eta_{ij}\text{Gamma}[\gamma_{ij}(t), 1], \\
\end{aligned}
$$
where $\mathbf{c}_{i}(t) = [c_{i1}(t), \dots, c_{iJ}(t)]$ is a latent reparameterization variable for $\boldsymbol{\psi}_i(t)$, $\delta_0$ is a Dirac measure, or point mass, at 0, $\eta_{ij} \sim \text{Bernoulli}(\theta_{j})$ is an at-risk indicator for taxon $j$ and individual $i$, $\theta_j \sim \text{Beta}(\alpha, \beta)$ is the taxon-specific probability of an at-risk zero, $T_{i}(t) = \sum_{j=1}^J c_{ij}(t)$, and $\gamma_{ij}(t)$ is the concentration parameter for taxon $j$ and individual $i$. The concentration parameters, $\gamma_{ij}(t)$, control the location of and variation around the mode of the relative abundances, $\boldsymbol{\psi}_i(t)$, on the simplex. \textcolor{black}{Since $\boldsymbol{\psi}_i(t)$ is a zero-inflated gamma variable normalized by the sum of the zero-inflated gamma variables all with the same rate parameter, the rate parameters cancel out and have no impact on the resulting zero-inflated Dirichlet distribution.} Observed zero counts can either be at-risk (i.e., $z_{ij}(t) = 0$ and $\eta_{ij} = 1$) or structural (i.e., $z_{ij}(t) = 0$ and $\eta_{ij} = 0$). At-risk zeros are zero counts that have a non-zero probability of occurrence; whereas, structural zeros are zeros that truly have a zero probability of occurrence. We assume the at-risk indicators, $\eta_{ij}$, do not depend on time, as zeros are only potentially structure if all an individual's counts are zero. \textcolor{black}{We assume the at-risk probability for a given taxon is shared across all individuals. We assign a common beta prior to each taxon-specific at-risk probability with hyperparameters $\alpha$ and $\beta$, effectively shrinking the at-risk probabilities towards the prior mean. While more complex relations between the at-risk probabilities may exist, we made this assumption since the observed zero counts provide no inherent information on whether they are structural or not. When prior information regarding the at-risk probabilities is available, such as established estimates of the proportion of at-risk observations, the hyperparameters $\alpha$ and $\beta$ may be specified to reflect this information. Alternatively, if there is a hypothesized relation between covariates and at-risk status, a regression framework linking $\theta_j$ to relevant covariates may be employed, similar to \cite{koslovsky2023bayesian}.}

To model the relation between a set of $P$ functional covariates and the taxa counts, we specify a log-linear regression model for the concentration parameters, $\gamma_{ij}(t)$. We define the taxon-specific log-linear regression model for the $j\sth$ concentration parameter as
\begin{equation}
\label{eq:gamma}
    \log[\gamma_{ij}(t)] = \beta_{j0}(t) + \sum_{p=1}^P\beta_{jp}(t)x_{ip}(t)+r_{ij},
\end{equation}
where  $\beta_{j1}(t), \dots, \beta_{jP}(t)$ are smooth functions that allow for the effect of the corresponding covariate to vary over time, $\beta_{j0}(t)$ is a population-level taxon-specific functional intercept term, and $r_{ij}$ is an intercept term specific to each individual-taxon combination. \textcolor{black}{The individual- and taxon-specific effects, $r_{ij}$, accommodate the repeated measures structure of the data by allowing for between-individual variation for each taxon and account for correlation among repeated measures for each individual.} 

We model the functional parameters with a B-spline basis expansion. Specifically, we assume $\beta_{jp}(t) = \mathbf{b}(t)\boldsymbol{\beta}^*_{jp}$. We define the basis vector $\mathbf{b}(t) = \left[1, b_1(t), \cdots, b_D(t)\right]$, where $b_1(t), \cdots, b_D(t)$ are generated using cubic B-splines with $D$ degrees of freedom without an intercept, and the corresponding vector of regression coefficients is $\boldsymbol{\beta}^*_{jp} = \left[\beta^*_{jp0}, \beta^*_{jp1}, \cdots, \beta^*_{jpD}\right]^\prime$. Including a 1 in $\mathbf{b}(t)$ allows $\beta_{jp}(t) = \beta^*_{jp0}$ when $\boldsymbol{\beta}^*_{jp} = \left[\beta^*_{jp0}, 0, \cdots, 0\right]^\prime$. We assume $\beta^*_{j00} \sim \text{Normal}(0, 1)$. To induce shrinkage for the regression coefficients of the smooth functions, we assume a global-local prior for $\beta^*_{jpd}$, $\forall p,d\text{ such that } (p,d) \neq(0,0)$, using the regularized horseshoe prior \citep{piironen2017sparsity}. That is, we assume $\beta^*_{jpd} \sim \text{Normal}(0,\sigma_{jp}^2)$, $\forall p,d\text{ such that } (p,d) \neq(0,0)$, where $\sigma^2_{jp} = \frac{\kappa_j^2\lambda_{jp}^{2}\tau_j^{2}}{\kappa_j^2 + \lambda_{jp}^{2}\tau_j^{2}}$, $\lambda_{jp}, \tau_{j} \sim \text{Half-Cauchy}(0,1)$, and $\kappa_j^{-2} \sim \text{Gamma}(\zeta, \rho)$. This allows non-active coefficients to shrink towards zero and effectively regularize the functional relations\textcolor{black}{, while controlling for potentially diverging variance that occurs with the horseshoe prior in weakly identifiable settings \citep{piironen2017sparsity}. The hyperparameters on the prior for $\kappa$ ($\zeta$ and $\rho$) can be chosen to be more or less informative on the level of shrinkage (see Supplement Section 3.1 for additional discussion).} To complete the model formulation, we assume $r_{ij} \sim \text{Normal}(0, \phi^2_j)$, where $\phi^{-2}_j \sim \text{Gamma}(a, b)$. 

The above model assumes all coefficients and covariates are functional. For covariates that are constant over time, the model immediately applies with $x_{ip}(t)=x_{ip}$ $\forall t\in\mathcal{T}$. For associations that are assumed to be constant over time, we model $\beta_{jp}(t) = \beta_{jp}$ and assume $\beta_{jp} \sim \text{Normal}(0,\sigma_{jp}^2)$. Furthermore, the model could easily extend to individual- and taxon-specific covariate effects, if applicable. 

\subsection{Inference}\label{sec:methInf}

Our primary interest is understanding how covariates may affect relative abundances and diversity over time. From the proposed model, we can calculate the relative abundance at time $t$ for taxon $j$ as a function of the model parameters:
\begin{equation}
\label{eq:RA}
    \RA = \frac{\exp[\beta_{j0}(t)+\mathbf{x}(t)\boldsymbol{\beta}_{j}(t)]}{\sum_{k=1}^J\exp[\beta_{k0}(t)+\mathbf{x}(t)\boldsymbol{\beta}_{k}(t)]},
\end{equation}
where $\boldsymbol{\beta}_{j}(t) = \left[\beta_{j1}(t), \beta_{j2}(t), \cdots, \beta_{jP}(t)\right]$ and $\mathbf{x}(t) = \left[x_{1}(t), x_{2}(t), \cdots, x_{P}(t)\right]'$ is a covariate profile vector. With scaled and centered continuous covariates, $\RA[j][t][\mathbf{0}]$ is interpreted as the expected relative abundance for the $j\sth$ taxon at time $t$ for individuals with the mean or reference categories of the covariates. The multiplicative difference in the relative abundance $j\sth$ taxon given a $v$-unit difference in the $p\sth$ covariate at time $t$ is
\begin{equation}
\label{eq:cRA}
    \chRA = \exp\left[v\beta_{jp}(t)\right]\frac{\sum_{k=1}^J \exp\left[\mathbf{x}(t)\boldsymbol{\beta}_{k}(t)\right]}{\sum_{k=1}^J \exp\left[\mathbf{x}(t)\boldsymbol{\beta}_{k}(t) + v\beta_{kp}(t)\right]}.
\end{equation}
Since the relative abundance of all taxa must sum to one, a change in the relative abundance of one taxon  results in a change in the relative abundance of at least one other taxon. \textcolor{black}{Additionally, since the multiplicative difference in relative abundance depends on all taxa's regression coefficients, the significance of a covariates's effect on a taxon's concentration parameter does not imply significance for that taxon's multiplicative difference in relative abundance.}

Researchers are also often interested in the $\alpha$-diversity of a composition, which is the diversity of the microbiome within an individual. We use Hill's diversity metric \citep{hill1973diversity} as defined in  \cite{roswell2021conceptual},
\begin{equation}
\label{eq:hillDiv}
    \Div = \left(\sum_{j=1}^J \RAin\left[\RAin\right]^{-l}\right)^{1/l},
\end{equation}
due to its flexibility in measuring compositions. The parameter $l$ controls the weight more abundant categories have on the metric. When $l = 1$, the count diversity is obtained; conversely, as $l \rightarrow 0$, $\Div$ approaches the Shannon diversity \citep{shannon1948mathematical,roswell2021conceptual}. Thus, when $l$ is closer to 1, categories are more evenly weighted, whereas $l$ closer to 0 weights the larger categories more. However, with our model, $\RA > 0\ \forall t,\ \mathbf{x}(t),\ j$, thus setting $l = 1$ always results in $\Div = J$. So, we recommend using $l \in [0, 1)$. We can examine how covariates affect the $\alpha$-diversity of microbial composition over time, providing a holistic view of the change. The multiplicative difference in $\alpha$-diversity with a $v$-unit difference in a covariate is
\begin{equation}
\label{eq:cDiv}
    \chDiv = \frac{\Div[p][t][\mathbf{x}(t)+\mathbf{\bar v}]}{\Div[p]},
\end{equation}
where $\mathbf{\bar v} = [0, \dots, 0, v, 0, \dots 0]$ provides a $v$-unit increase to the covariate $p$ of $\mathbf{x}(t)$.

\subsection{Posterior Sampling}

To sample from the posterior distribution, we implement a Metropolis-Hastings (MH) within Gibbs algorithm\textcolor{black}{, which is fully detailed in Section 1 of the Supplementary Material}. \textcolor{black}{Briefly, to efficiently sample the horseshoe prior parameters, we apply the auxiliary parameterization proposed by \cite{makalic2015simple}. Also, we integrate out $\theta_j$ to efficiently sample $\eta_{ij}$, which results in a beta-binomial prior.} The parameters $\eta_{ij}$, $\beta^*_{jpd}$, and $r_{ij}$ are sampled using MH updates \textcolor{black}{described in Sections 1.1 and 1.2 of the Supplementary Material}, while the rest of the parameters are sampled using Gibbs updates \textcolor{black}{described in Section 1.3 of the Supplementary Material}. Lastly, we incorporate a proposal adjustment algorithm for $\boldsymbol{\beta}^*_{jp}$ during burn-in to allow for efficient sampling \textcolor{black}{described in Section 1.4 of the Supplementary Material}. The inherent dependence among parameters can cause the samples to drift around the mean rather than sample stably around it. While we did not find that this was an issue in our analysis, we recommend running multiple chains and thinning the MCMC samples to help assess convergence. 

\section{Analysis of Infant Microbiome} \label{sec:case}

Using the preterm infant microbial samples collected by \cite{la2014patterned}, \textcolor{black}{we investigated the associations between percentage of breast milk in the diet and gestational age at birth and the 16 taxa over time}, controlling for the other covariates, with the proposed model. Our analysis provides inference on the multiplicative difference in relative abundance, $\chRA$, and the multiplicative difference in $\alpha$-diversity, $\chDiv$.

\subsection{Data and Model Setup}\label{sec:caseSetup}

To ensure balanced shrinkage across covariates, the proportion of observed days with antibiotic use and gestational age at birth were centered and scaled across all observations. Additionally, we included a binary variable to indicate whether a sample was taken pre or post a change in the sampling procedure, following \cite{la2014patterned}. We allow this variable to have a constant coefficient (i.e., $\beta_{jp}(t) = \beta_{jp}$). We set breast milk percentage $<$$10\%$, female, c-section, and multi-patient room as the reference categories. This makes the baseline relative abundance at time $t$, $\RA[j][t][\mathbf{0}]$, interpreted as the effect of time on the $j\sth$ relative abundance for infants that are female, in a multi-patient room, delivered by c-section, had an average proportion of total days on antibiotics, were 27 weeks of gestational age at birth, and were fed $<$$10\%$ breast milk. Additionally, we included individual- and taxon-specific intercept terms for each infant. 

\textcolor{black}{We specified the hyperparameters using a combination of weakly informative and informative priors. We set the individual- and taxon-specific intercepts, $r_{ij}$, variance hyperparameters to $a=3$ and $b=9$, implying a mean variance of three. We set the at-risk indicator, $\eta_{ij}$, hyperparameters to $\alpha=0.01$ and $\beta=10$, representing a small probability of at-risk observations. Making the at-risk indicator probabilities informative is necessary since the observed zero counts carry no information on whether they are at-risk or structural. Furthermore, the at-risk indicator can only have a positive probability of equaling zero if all counts for an individual were zero. This assumes that if all observed counts for a taxon were zero for an individual, then the observed zeros were structural zeros with high probability; otherwise, they were considered at-risk. We set the hyperparameters for the regularized horseshoe parameter $\kappa$ to $\zeta=100$ and $\rho = 900$. These are set to protect against over-shrinkage for taxa with high amounts of zero-inflation and active parameters. Since our objective in using the regularized horseshoe is to mitigate divergence for highly zero-inflated taxa with active functional covariates, having an informative prior with a large mean protects against divergence while inducing smaller amounts of shrinkage. Lastly, we set the dimension of the B-spline basis for the functional coefficients to $D=4$ to have smooth functional coefficients.}

We ran four MCMC chains for 85,000 iterations with \textcolor{black}{45,000} iterations treated as burn-in, thinning to every \textcolor{black}{$40\sth$ iteration due to memory constraints. With an average runtime of 1 hour and 5 minutes across the four chains on a single core of an AMD Milan x86-64 CPU, performing this number of iterations to ensure chain convergence was feasible. We assessed convergence with thinned trace plots of the multiplicative difference in relative abundances (Section 2.2 of the Supplementary Material).} Inferential results were based on the remaining 4,000 samples. \textcolor{black}{We initialized most of the parameters by sampling from their prior specifications, except the regression coefficients, $\boldsymbol{\beta}^*_{jp}$, individual- and taxon-specific intercept terms, $r_{ij}$, and at-risk indicators, $\eta_{ij}$. We initialized the regression coefficients and individual- and taxon-specific intercept terms by sampling from a Uniform(-.75, .75) and Uniform(-.05, .05), respectively. We initialized the at-risk indicators to zero if all counts for a given taxon were zero across time for an individual, and one otherwise.} 

We focused our inference on the three most abundant taxa: Bacilli, Clostridia, and Gammaproteobacteria. Although we modeled the entire sample period, we restricted inference to the first 50 days of life to ensure it was representative of all infants in the study, as days 50 to 80 only contain the infants with less than 28 weeks of gestational age at birth, as seen in Figure \ref{fig:idvInfandGARA}. For the multiplicative difference in $\alpha$-diversity, we set $l=.75$ due to the dimension of the data and most observed infant compositions being dominated by the three taxa.

\subsection{Results}

\begin{figure}[h!]
    \centering
    \includegraphics[width=1\textwidth]{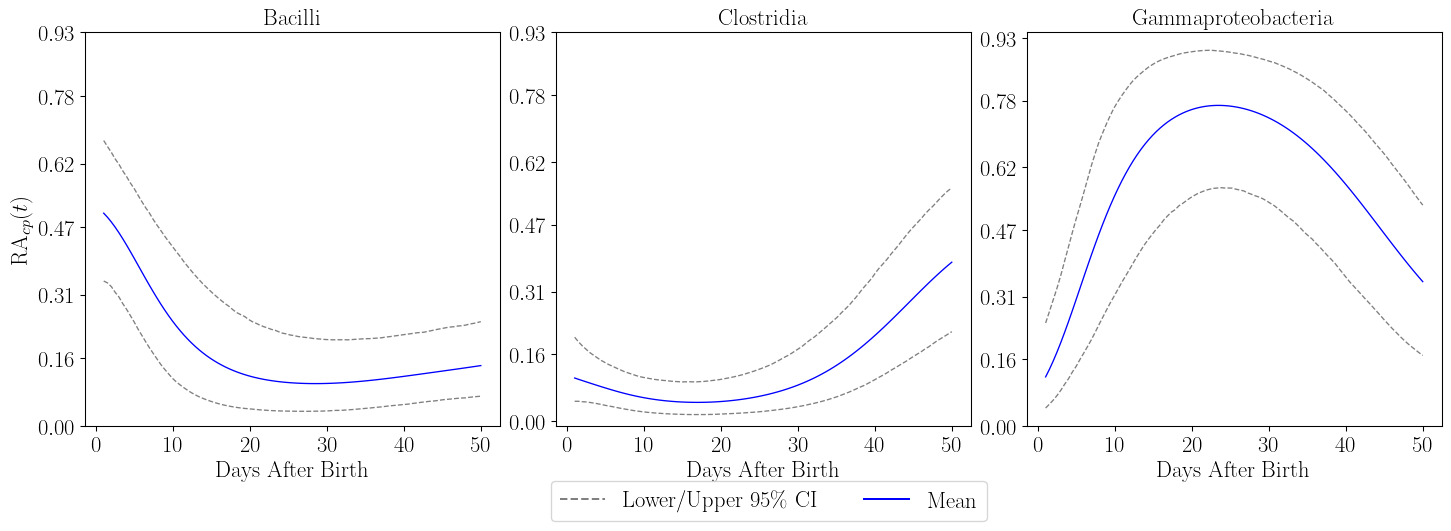}
    \caption{Estimated difference in relative abundance of Bacilli, Clostridia, and Gammaproteobacteria over time as baseline covariate values ($\mathbf{x}(t)=\mathbf{0}$). The figure shows the posterior mean (solid blue line) and 0.95 probability credible intervals (dashed black lines). Bacilli, Clostridia, and Gammaproteobacteria were the three most abundant taxa at baseline values of the covariates.}
    \label{fig:interRAs}
\end{figure}

Figure \ref{fig:interRAs} shows the baseline relative abundance for Bacilli, Clostridia, and Gammaproteobacteria over time. We observed a decrease in relative abundance for Bacilli over time, becoming stable after the first 20 days; Clostridia's relative abundance increased after about 30 days; and Gammaproteobacteria's relative abundance increased over the first 15 days, then decreased after 30 days. For most other taxa, we found a constant intercept term (i.e., $\beta_{j0}(t) \approx \beta^*_{j00}$) and a decrease in relative abundance over the first 20 days of life, after which they became relatively stable.

\begin{figure}[h!]
    \centering
    \begin{subfigure}[t]{0.595\textwidth}
      \centering
        \includegraphics[width=1\linewidth]{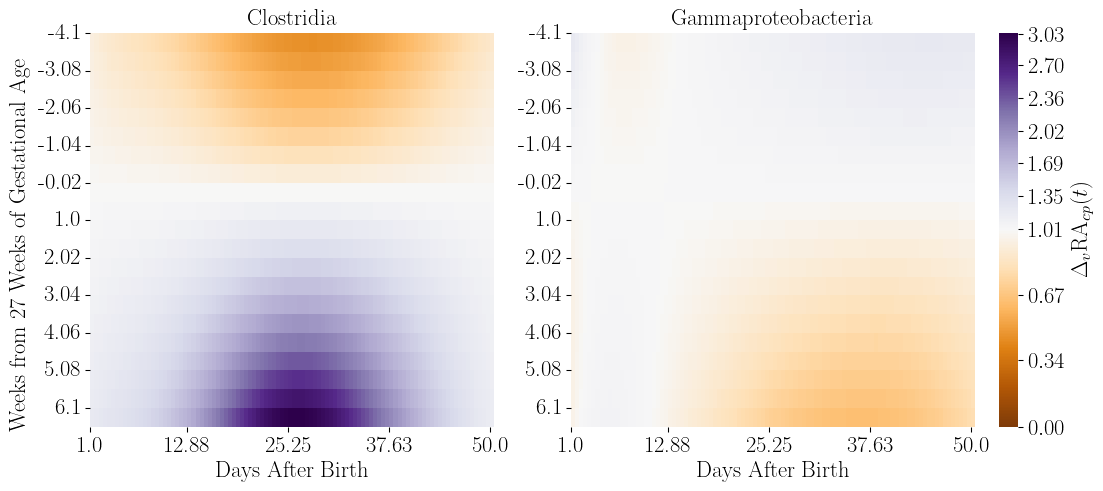} 
        \caption{Multiplicative Difference in Relative Abundance} \label{fig:GARA}
    \end{subfigure}
    \begin{subfigure}[t]{0.3855\textwidth}
      \centering
        \includegraphics[width=1\linewidth]{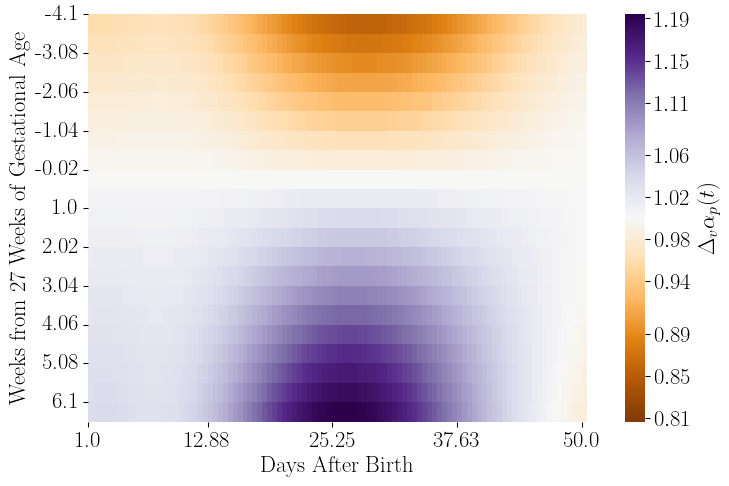} 
        \caption{Multiplicative Difference in $\alpha$-Diversity} \label{fig:GADiv}
    \end{subfigure}
    \caption{The estimated mean multiplicative difference in relative abundance, $\chRA$, and $\alpha$-diversity, $\chDiv$, with a $v$-weak difference in gestational age at birth for Clostridia and Gammaproteobacteria. The left plot shows the $\chRA$ for Clostridia and Gammaproteobacteria, and the right plot shows the $\chDiv$ with $l=0.75$.}
    \label{fig:GAs}
\end{figure}

We explored the multiplicative difference in relative abundance, $\chRA$, and $\alpha$-diversity, $\chDiv$, associated with a $v$-week difference in gestational age from baseline (27 weeks). Since age is a continuous variable, we visualized the mean multiplicative difference in relative abundance and $\alpha$-diversity using a heatmap in Figure \ref{fig:GAs}. For infants born at higher gestational ages, we estimated a higher average abundance of Clostridia, holding all else constant. In contrast, the multiplicative difference in relative abundance for Gammaproteobacteria was less than 1 for infants closer to full-term, implying less abundance for infants with higher gestational age at birth. The trend was reversed for more premature infants. \textcolor{black}{We did not observe} any associations between gestational age and Bacilli. We found that the $\alpha$-diversity \textcolor{black}{was lower} among the most premature infants and was higher for those closer to full term (Figure \ref{fig:GADiv}). In Supplementary Figure S6, we threshold the results presented in Figure \ref{fig:GAs} based on whether or not the point estimates' corresponding 0.95 probability credible intervals contained one for reference.  

\begin{figure}[h!]
    \centering
    \includegraphics[width=0.9\textwidth]{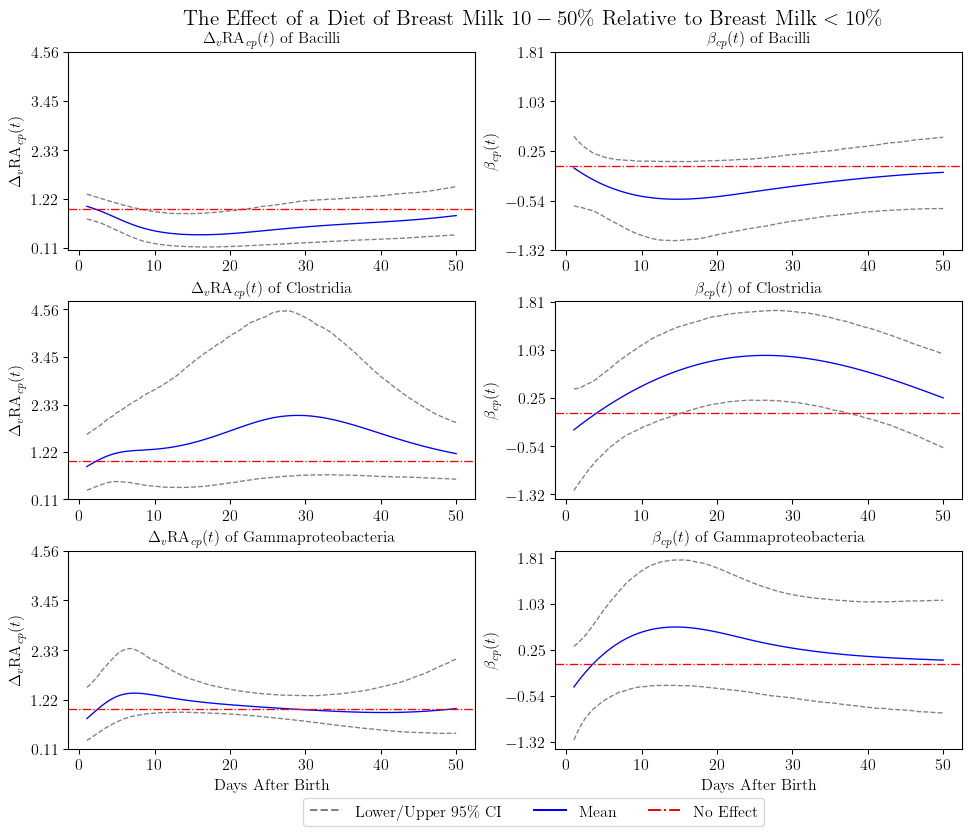}
    \caption{The estimated multiplicative difference in relative abundance, $\chRA$, for infants on a diet of 10-50\% breast milk compared to $<$$10\%$ breast milk for the three most abundant taxa (left plot). For reference, we provide the estimated functional coefficients, $\beta_{jp}(t)$, for infants on a diet of 10-50\% breast milk (right plot).}
    \label{fig:BMRA1050}
\end{figure}

We present the mean multiplicative differences in relative abundances for Bacilli, Clostridia, and Gammaproteobacteria with a diet of $10-50\%$ and $>50\%$ breast milk compared to $<$$10\%$ in Figure \ref{fig:BMRA1050} and Supplementary Figure S7, respectively. For a diet with a breast milk percentage of $10-50\%$ compared to $<$$10\%$, Bacilli's level of decrease in relative abundance grew until 10 days after birth, then slowly returned to no difference over 25 to 50 days after birth; Clostridia's level of increase in relative abundance rose until around 30 days after birth, then returned to no effect by 50 days after birth; and Gammaproteobacteria's relative abundance showed no differences over the assessment period. For a diet with a breast milk percentage of $>50\%$ compared to $<$$10\%$, we observed less evidence of a difference in relative abundance for these three taxa (Supplementary Figure S7). We observed an increase in the mean multiplicative difference in $\alpha$-diversity over the assessment period for a diet of $>50\%$ breast milk relative to $<$$10\%$ (Supplementary Figure S8). Diets with $10-50\%$ breast milk compared to $<$$10\%$ had a marginally lower $\alpha$-diversity over days 10 to 50 after birth. 

These findings are in line with many previous research studies on premature infants or breast milk in infant diets. \cite{chernikova2018premature} and \cite{healy2022clinical} reported a lower microbial diversity for preterm infants compared to full-term infants. Additionally, \cite{healy2022clinical} highlighted the dominance of Gammaproteobacteria in preterm infants, controlling for breast milk, mode of delivery, and antibiotics. For breast-fed infants, \cite{ma2020comparison} documented no difference in $\alpha$-diversity for the first 3 months of life, followed by a significant difference at 6 months of age, and \cite{cong2016gut} found that breast milk percentage in the diet was associated with higher $\alpha$-diversity. Both \cite{ma2020comparison} and \cite{gregory2016influence} identified Bacilli as more prevalent in formula-fed infants. While \cite{pannaraj2017association} associated higher Bacilli abundances with breast milk.

\textcolor{black}{Comparing our results to the original analysis, we found similar trajectories for the three main taxa's relative abundances over time \citep{la2014patterned}. However, we did not find the same relations between the percentage of breast milk in the diet and the main taxa. \cite{la2014patterned} fit a mixed model for Bacilli, Clostridia, and Gammaproteobacteria's relative abundances individually and found that Gammaproteobacteria was positively and Bacilli and Clostridia were negatively associated with percentage of breast milk in the diet. Whereas, our analysis suggested Gammaproteobacteria was not associated, Clostridia positively associated, and Bacilli negatively associated with the percentage of breast milk in the diet over the duration of the study. Our model further identified non-constant effects for Clostridia and Bacilli and a positive association between $\alpha$-diversity and the percentage of breast milk in the diet. One of the strengths of our analysis is that we accommodate the compositional structure of the data, which can lead to spurious associations if ignored \citep{gloor2017microbiome}. Using our proposed method, we were able to gain additional insights into the effects over time without grouping and provide inference on relative abundances and $\alpha$-diversity with a single model. Our model provides clinicians insights on potentially effective time periods and subgroups of infants to recommend for dietary intervention. Follow-up explanatory studies could enrich the understanding of these findings and help design heterogeneous treatment regimes.}

\textcolor{black}{In Supplementary Section 2.1, we present a sensitivity analysis for the case study to further justify the assumed prior specifications. We found the results were robust to most prior choices for breast milk percentage in the diet and gestational age of the infant. In Supplementary Section 2.4, we present posterior predictive checks to assess model fit. The posterior predictive checks indicate adequate model fit for the covariance between taxa and the relative abundances of the taxa.}

\section{Simulation Study}\label{sec:sim}

We implemented a simulation study to demonstrate the scalability and estimation performance of the proposed model in settings similar to those observed in microbiome research. We considered two simulation scenarios. First, we considered a scenario with 50 taxa and compared the proposed method's performance to alternative methods. Second, we demonstrated the scalability of the proposed method in scenarios with $J = 50,\ 250,\ 500,$ and $1000$ taxa. 

\subsection{Data Generation}

For both scenarios, the data generation procedure was the same other than the number of taxa. \textcolor{black}{We fixed the number of individuals to 50 and} generated the number of observations per individual from a Discrete-Uniform(3, 10) and the observation time points from a Uniform(0, 10). We generated $P = 10$ covariates at each sample time point from a Normal(0,1) for each individual. For the functional coefficients, we set 
$$
\begin{aligned}
    f_1(t) &= [-0.2\left(t-5\right)^{2}+5]/7, \\
    f_2(t) &= \frac{1}{[1.75+e^{-1.25\left(t-5\right)}] }, \\
    f_3(t) &= 0.07t, \\
    f_4(t) &= 0.5,\text{ and }\\
    f_5(t) &= 0,
\end{aligned}
$$
where $f_5(t)$ corresponds to no effect (see Supplementary Figure S12 for a visualization of the figures). For each of the intercept terms, $\beta_{j0}(t)$, we randomly assigned $\pm f_k(t)$ where $k=1,2,3,4$ with equal probability. For covariates $p=1,\dots, 10$ and taxa $j=1, 2, 3, 4$, we randomly assigned $\beta_{jp}(t) = \pm f_k(t)$, where $k$ was chosen from $1,\dots,5$  with equal probability and $\beta_{jp}(t) = f_5(t)$ for $j=5, 6,\dots, 50$. Hence, there was only a direct effect of covariates on the first four taxa\textcolor{black}{, with a multiplicative effect on the concentration parameters ranging from 0.5 to 2 with a one-unit increase in the corresponding covariate.} We sampled the individual- and taxon-specific intercepts from a Uniform(-.05, .05) and $\theta_j$ from a mixture of three different uniform distributions with equal probability to control the proportion of zeros in the data (Uniform(0, .15), Uniform(.15, .75), or Uniform(.75, .90)). Then, we sampled each individual's at-risk indicators, $\eta_{ij}$, from a Bernoulli($\theta_j$). We resampled the at-risk indicators if there were fewer than 5 individuals with $\eta_{ij} = 1$ to ensure we did not have too few functional observations. For the second scenario, which evaluates scalability, we implemented the same design with the effects of taxa $51,\dots, J$ set to $f_5(t)$, so that they did not affect the composition.

For the first scenario with $J = 50$ taxa, we compared our model with variations of a DM model to allow for direct comparison of the results. The first model we compared to is a standard DM regression model, which ignores potential zero-inflation. We assumed the same functional coefficient structure with B-splines and regularized horseshoe priors as in the proposed model. However, this model does not contain any individual- and taxon-specific effects. We refer to this model as DM. The second method, which we refer to as ZIDM, has the same zero-inflated DM structure and functional coefficient structure as the proposed model, but no individual- and taxon-specific effects. This model is similar to that of \cite{koslovsky2023bayesian} but uses regularized horseshoe priors in place of spike-and-slab priors to model the functional coefficients. Lastly, we consider a model that is similar to the model introduced in \cite{pedone2023subject}, which provides subject-specific regression coefficients by allowing all covariates to have a coefficient comprised of a constant component (main effect) and a functional component based on the other covariates (effect modifiers). If we only allow time and individual id to be effect modifiers, then we obtain a similar model to our approach without zero-inflation. To ensure comparable results, we implemented this version of the model ourselves instead of relying on their available software. We refer to this model as FunC-DM. \textcolor{black}{For all models, we set the hyperparameters and parameter initializations similar to that described in Section \ref{sec:case}. In Supplementary Section 3.1, we provide a sensitivity analysis to assess model sensitivity to hyperparameter specification. We find our model is sensitive to the zero-inflation proportion hyperparameters and fairly robust to the other prior specifications. Additionally, in Supplementary Section 3.3, we compare the performance of all of the methods under model misspecification. We find that the proposed method obtained the best estimation performance in this scenario.}

To evaluate the fit of the models, we computed statistics based on the sampled functional coefficients, $\beta_{jp}(t)$, and the multiplicative difference in relative abundance, $\chRA[1][j]$. Specifically, we calculated the coverage rate of the pointwise 0.95 credible interval of true functional coefficients, $\pm f_k(t)$, and multiplicative difference in relative abundance; average root mean squared error (ARMSE) of the multiplicative difference in relative abundance; and mean Aitchison distance \citep{aitchison1992criteria} between the estimated and true relative abundances. We evaluated the credible intervals and squared errors averaged over 100 equally-spaced time points from 0 to 10. Specifically, we set
\begin{equation}
\begin{aligned}
\label{eq:eval}
\text{RA95} &= \frac{1}{100}\sum_{t\in T} I\left\{ Q_{0.025}\left(\left[\chRAin[1][j]^{(s)} \right]_{s=1}^{S}\right) \le \omega_{jp}(t) \le Q_{0.975}\left(\left[\chRAin[1][j]^{(s)} \right]_{s=1}^{S}\right)\right\},\\
\text{B95} &= \frac{1}{100}\sum_{t\in T} I\left( Q_{0.025}\left[\left\{\beta_{jp}(t)^{(s)} \right\}_{s=1}^{S}\right] \le \pm f_k(t) \le Q_{0.975}\left[\left\{\beta_{jp}(t)^{(s)} \right\}_{s=1}^{S}\right]\right),\text{ and }\\
\text{ARMSE} &= \frac{1}{S}\sum_{s=1}^S \sqrt{\frac{1}{100}\sum_{t\in\mathrm{T}}\left[\omega_{jp}(t) - \chRAin[1][j]^{(s)}\right]^2},
\end{aligned}
\end{equation}
where $Q_q$ is the quantile function, $I(\cdot)$ is the indicator function, $\omega_{jp}(t)$ is the true multiplicative difference in relative abundance, $\{\cdot \}_{s=1}^{S}$ and $[\cdot ]_{s=1}^{S}$ are the set of samples, $S$ is the number of MCMC samples, and $T = \{0, .1, \dots, 10\}$. Lastly, the Aitchison distance was evaluated at each observation, and we refer to the mean as MeAD.

\subsection{Simulation Results}

We present the results of scenario 1 for the proposed and alternative models in Table \ref{tab:multiPResults}. For the ZIDM model, 8 of the 100 simulation seeds did not converge and were removed from the results; the other models converged across all seeds. For ARMSE, FunC-ZIDM and FunC-DM performed the best overall for taxa with active functional covariates. For interval coverage of $\chRA[1][j]$ and $\beta_{jp}(t)$, FunC-ZIDM performed significantly better than the alternative models for taxa with active functional covariates with roughly 95\% coverage. The other three methods had coverage ranging from  60\% to  87\%. FunC-DM and DM obtained weaker performance than their zero-inflated counterparts because ignoring zero-inflation biased coefficient estimates towards 0 and constant functions over time due to the extra zeros used in the estimate. Additionally, we observed that ignoring zero-inflation lowered uncertainty due to the number of zeros included in the parameter estimation. This explains why FunC-ZIDM performed much better in terms of interval coverage for $\chRA[1][j]$ and $\beta_{jp}(t)$ but similarly compared to FunC-DM in terms of ARMSE for taxa with active functional covariates. Due to the induced shrinkage for taxa with non-active functional covariates, all models obtained coverage above the nominal level for both $\chRA[1][j]$ and $\beta_{jp}(t)$. Additionally, all models demonstrated very low ARMSE for these taxa, although FunC-ZIDM was slightly higher than the others. FunC-ZIDM performed the best in terms of MeAD.

\begin{table}[h!]
\centering
\begin{tabular}{l r rrr rrr}
\toprule
 &  & \multicolumn{3}{c}{Active Coefficients} & \multicolumn{3}{c}{Non-active Coefficients} \\
 &  & \multicolumn{3}{c}{$(j \in \{1, \dots,4\})$} & \multicolumn{3}{c}{$(j \in \{5, \dots,50\})$} \\
 \cmidrule(lr){3-5} \cmidrule(lr){6-8}
 &  MeAD & RA95 & B95 & ARMSE & RA95 & B95 &  ARMSE \\
Model & & & & & & & \\
\midrule
FunC-ZIDM  & \textbf{8.3751} & \textbf{0.9356} & \textbf{0.9348} & \textbf{0.1903} & \textbf{0.9981} & 0.9987 & 0.0461\\
ZIDM & 27.6850 & 0.8711 & 0.8703 & 0.2353 & 0.9949 & \textbf{0.9998} & 0.0373 \\
FunC-DM    & 27.2143 & 0.7663 & 0.7649 & 0.1950 & 0.9836 & 0.9993 & \textbf{0.0326} \\
DM   & 29.5917 & 0.6076 & 0.6099 & 0.2579 & 0.9811 & 0.9983 & 0.0369 \\
\bottomrule
\end{tabular}
\caption{Simulation results for scenario 1. The table shows 95\% CI coverage rate for $\chRA$ (RA95) and $\beta_{jp}(t)$ (B95), average root mean squared error (ARMSE), and mean Aitchison distance (MeAD) for each model. Taxa 1-4 had active coefficients, while 5-50 were inactive. The best of each metric is bolded.}
\label{tab:multiPResults}
\end{table}

\begin{figure}[h!]
    \centering
    \includegraphics[width=0.9\textwidth]{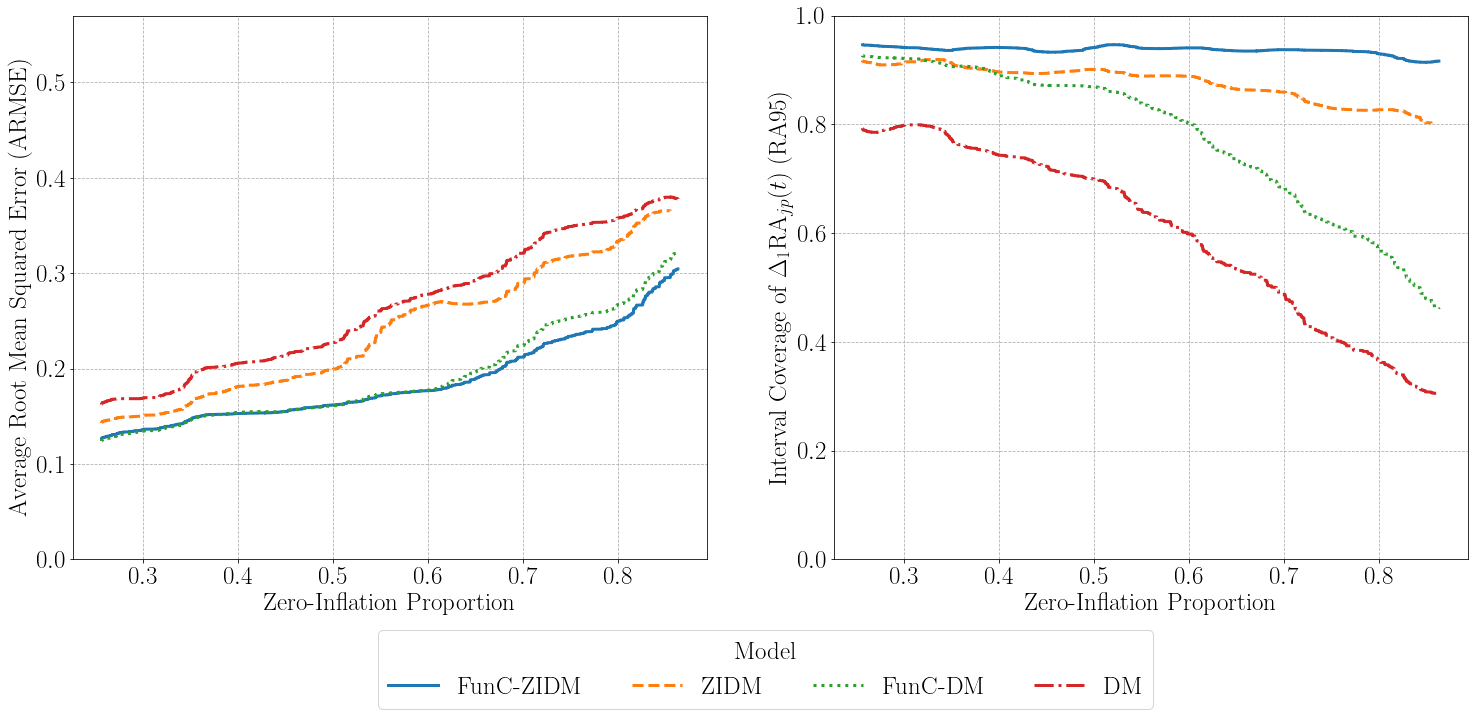}
    \caption{Results from simulation scenario 1 demonstrating performance as a function of zero-inflation level. The figure shows Gaussian kernel moving averages with a bandwidth of 750 across levels of zero-inflation for the active covariates. Each line represents a different model.}
    \label{fig:compet}
\end{figure}

Figure \ref{fig:compet} shows how performance changed as a function of zero-inflation level for the taxa with active functional covariates. We found that the ARMSE increased for all models, with the proposed model and FunC-DM having similar ARMSE over most zero-inflation levels. At higher levels of zero-inflation, FunC-ZIDM had slightly lower ARMSE than FunC-DM. For interval coverage of $\chRA[1][j]$, the proposed model performed better than the alternative models across all zero-inflation levels. This was especially pronounced at high levels of zero-inflation where the three alternative models had coverage rates well below the nominal level, while FunC-ZIDM had coverage near the nominal level for all zero-inflation levels. ARMSE performance of the proposed model was similar to FunC-DM, whereas interval coverage for $\chRA[1][j]$ improved, due to the proposed model estimating more parameter uncertainty as zero-inflation increased. 

\begin{figure}[h!]
    \centering
    \includegraphics[width=0.9\textwidth]{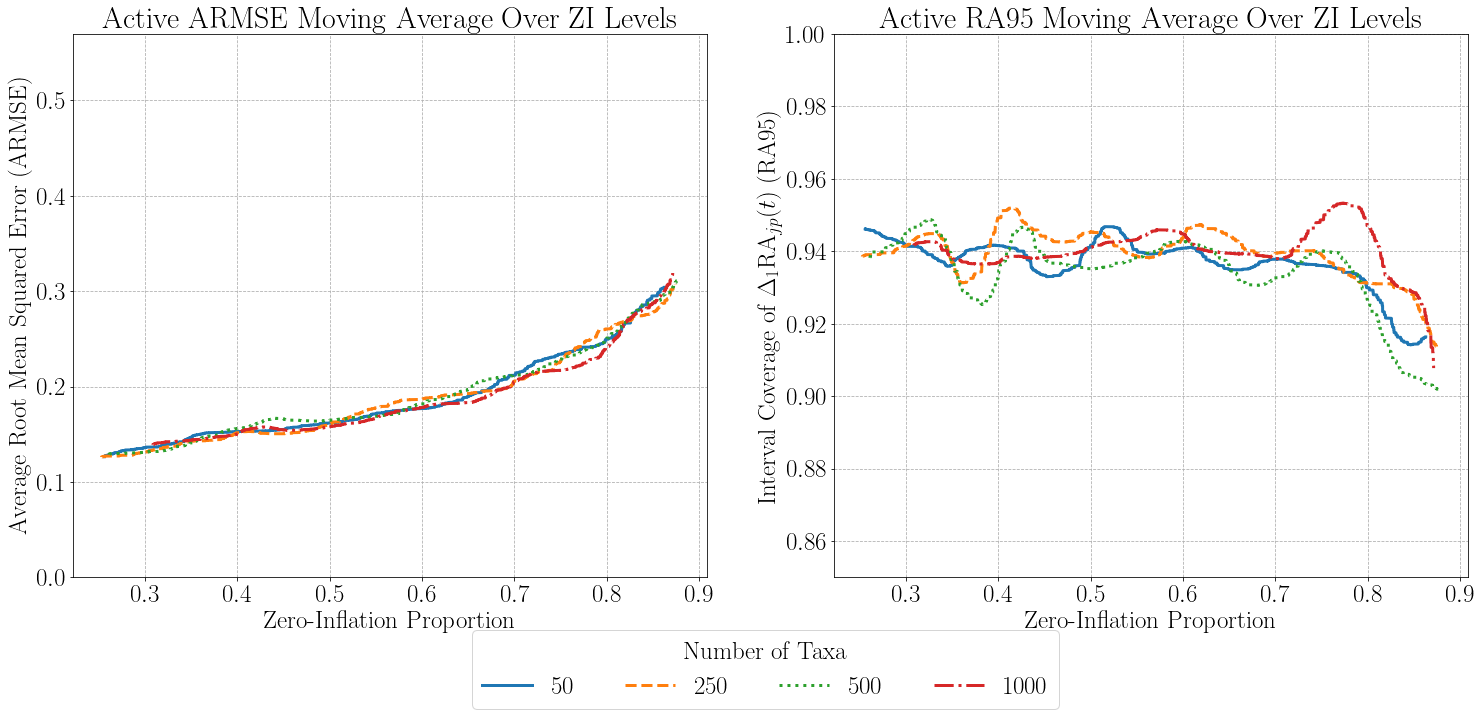}
    \caption{Simulation results for scenario 2. The figure shows Gaussian kernel moving averages with a bandwidth of 750 across levels of zero-inflation for the active covariates. Each line represents results for a different number of taxa in the simulated data.}
    \label{fig:scal}
\end{figure}

For scenario 2, we present the ARMSE and interval coverage for $\chRA[1][j]$ for the taxa with active functional covariates by level of zero-inflation for $J=50,\ 250,\ 500,$ and $1000$ taxa in Figure \ref{fig:scal}. Neither the ARMSE nor the coverage for $\chRA[1][j]$ differed as the number of taxa increased. These results demonstrate our method scales to the large number of categories often seen in microbiome studies without sacrificing the performance of the model. We show additional results for scenario 2 with 250 taxa that show performance separately for each functional coefficient in Supplementary Figure S13.

The run time for the proposed model scales with the number of categories, $J$, covariates, $P$, degrees of freedom, $D$, and observations, $\sum^N_{i=1}|\{t\}_i|$. The average run times were 10, 42, 84, and 194 minutes with $D=4$ and $P = 10$ for $J = 50,\ 250,\ 500,$ and $1000$, respectively, on a single core of an AMD Milan x86-64 CPU. 

\section{Discussion}

In this work, we propose a functional concurrent regression model that handles the compositional structure of microbiome data while accounting for zero-inflation, repeated measures, overdispersion, and time-varying effects. The approach uses a zero-inflated Dirichlet-multinomial model with additive functional covariates. We induce shrinkage through a regularized horseshoe prior on the spline coefficients and include random effects to account for between-individual variation. We provide an \texttt{R} package \texttt{FunCZIDM} and a \texttt{shiny} app to generate dynamic figures capturing the inferential results. 

Through application and simulation, we show the effectiveness of the model to capture trends over time and scale to large compositional spaces. We also show that the method maintains nominal interval coverage rates even when high zero-inflation occurs. We provide individual- and compositional-level inference through the multiplicative difference in relative abundance, $\chRA$, and the multiplicative difference in $\alpha$-diversity, $\chDiv$, for different covariate profiles. 

Our case study results have similar findings as others on the association of breast milk in the infant diet and gestational age at birth with specific taxa and diversity. Our model has the added benefit of smoothly-varying covariate effects over time, which capture the temporal dynamics of the microbiome known to change during infancy. Our findings suggest several taxa have a functional effect over time on relative abundances and $\alpha$-diversity. Further exploration of the findings can be conducted with the \texttt{R} package and associated \texttt{shiny} app. The absence of time-varying effects in inferential models could contribute to the lack of reproducibility across studies of microbiome data, as this could lead to averaging effects over the study period. This highlights the importance of methods like FunC-ZIDM that have the ability to capture the underlying functional effects, as reproducibility is essential in health-related research settings.

\textcolor{black}{The strictly negative correlation structure between categories of a DM distribution is a well-documented limitation. Many distributions have been developed to further account for positive correlations, including the Dirichlet-tree, generalized Dirichlet, and extended flexible Dirichlet distributions \citep{connor1969concepts,dennis1991hyper,ongaro2020new}. \cite{MENEZES2025105492} recently showed that the mixture formulation of the ZIDM model is able to accommodate both negative and positive correlations. Furthermore, the addition of functional trajectories and random effects in FunC-ZIDM adds even more flexibility. We demonstrate that the proposed model is able to capture both positive and negative covariance structure in the observed taxa counts empirically through posterior predictive checks (Figure S9 of the Supplementary Material).}

In this work, we assume covariate effects are time-varying; though, the proposed model could be used to model varying coefficients for any continuous covariate. For example, we could apply the model to investigate how the effect of milk consumption on taxa relative abundances varies as a function of gestational age at birth. Additionally, the inference we provide in this work and the associated \texttt{R} package is for a specific covariate profile. One may be interested in the effect of a covariate on average over the covariate space, effectively marginalizing out $\mathbf{x}(t)$. This could be done by averaging over the covariates in the data; however, a more robust way to estimate the distribution of $\mathbf{x}(t)$ is with the Bayesian bootstrap \citep{rubin1981boot}.

The proposed model is the first to allow for functional coefficients, to handle repeated measures, and to address the compositional structure in the presence of zero-inflation. With the emergence of larger longitudinal microbiome datasets, the proposed method allows for a more in-depth analysis while respecting compositionality and zero-inflation.

\section*{Supplementary Material}

Supplementary Material contains the details of the posterior sampler, additional figures and sensitivity analyses  for the simulation study and case study, convergence assessments and posterior predictive checks for the case study, and model misspecification results for the simulation study. The Supplementary Material is available at \textit{Biostatistics Journal} online.

\section*{Conflict of Interest}

None declared.

\section*{Code and Data Availability}

The \texttt{R} package \texttt{FunCZIDM} implementing the proposed method can be found at \url{https://github.com/brodyee/FunCZIDM}. The package also contains scripts used to run the simulations and case study and the data used in the case study.

\section*{Funding}

This work was supported by the National Science Foundation grant DMS-2245492 and the National Institutes of Health grant R01ES035735. The content is solely the responsibility of the authors and does not necessarily represent the official views of the National Science Foundation or National Institutes of Health.

\bibliographystyle{apalike}
\bibliography{references}

\end{document}